\documentclass[epj]{webofc}
\usepackage[varg]{txfonts}   % Web of Conferences font
\woctitle{MESON2014 - the 13$^\textrm{th}$ International Workshop on Meson Production, Properties and Interaction}
\begin{document}
\selectlanguage{english}
\title{Exotic and conventional mesons from lattice}

% insert email only for speaker/presenter
\author{Sasa Prelovsek\inst{1,2}\fnsep\thanks{\email{sasa.prelovsek@ijs.si}}  }

\institute{ Faculty of Mathematics and Physics, University of Ljubljana, Jadranska 19, Ljubljana, Slovenia\and
        Jozef Stefan Institute, Jamova 39, Ljubljana, Slovenia       }

\abstract{%
Recent results on meson spectroscopy from lattice QCD are reviewed.  The emphasis is on interesting states near thresholds like   $Z_c^+$, $X(3872)$ and $D_s^0(2317)$. Another focus is on the meson resonances in light, strange and charm sector, where  the resonance masses as well as the strong decay widths are extracted from the lattice.   }
\maketitle
\section{Introduction}

Lattice QCD aims at describing hadrons and interactions between them based on first-principle Quantum ChromoDynamics. It is based on the evaluation of correlation functions by means of Feynman path integral of QCD action on discretized and finite Euclidean space-time. I review recent results on spectrum  for the exotic and conventional mesons.  

 \section{ The discrete spectrum from lattice and information encoded in it}

The physics information on a meson (below, near or above threshold) is commonly extracted from the discrete energy spectrum in lattice QCD. 
The physical system for given quantum numbers is created from the vacuum $|\Omega\rangle$ using interpolator ${\cal O}_j^\dagger$  at time $t\!=\!0$ and the system propagates for time $t$ before being annihilated by ${\cal O}_i$.    To study a meson state with given $J^P$ one  can use ${\cal O}\simeq \bar q	\Gamma q,~$ $(\bar q \Gamma_1  q)_{\vec p_1}(\bar q \Gamma_2  q)_{\vec p_2},~$$[\bar q \Gamma_1 \bar q][q\Gamma_2 q]$ with desired quantum numbers.  After the spectral decomposition the correlators are expressed in terms of the energies  $E_n$ of eigenstates $|n\rangle$ and their overlaps $Z_j^n$
\begin{equation}
\label{C}
C_{ij}(t)= \langle \Omega|{\cal O}_i (t) {\cal O}_j^\dagger (0)|\Omega \rangle=\sum_{n}Z_i^nZ_j^{n*}~e^{-E_n t}~,\qquad Z_j^n\equiv \langle \Omega|{\cal O}_j|n\rangle~.
\end{equation}
The most widely used method to extract the discrete spectrum $E_n$ and overlaps  from the correlation matrix $C_{ij}(t)$ is the generalised
eigenvalue method  $C(t)u^{(n)}(t)=\lambda^{(n)}(t)C(t_0)u^{(n)}(t)$ \cite{Michael:1985ne,Blossier:2009kd}.
The energies $E_n$ are extracted from the  exponential
behavior of the eigenvalues $\lambda^{(n)}(t)\propto  e^{-E_n t}$ at large $t$.   

 All physical eigenstates with given quantum numbers appear as energy levels in principle. These can be  "one-meson" states (for example $\chi_{c1}$ in $1^{++}$ charmonium channel), "two-meson" states (for example $D\bar D^*$) and the multi-meson states  (for example $J/\psi \pi\pi$).  In reality the eigenstates are mixtures of these Fock components. Three- and more-meson states have never been taken into account in the study of the meson spectroscopy yet. The major step during the preceding years came from treating two-meson states rigorously. 
 These  have discrete spectrum due to periodic boundary condition on finite lattice.  If the two mesons do not interact, then momenta of each meson is $\vec{p}= \!\tfrac{2\pi}{L}\vec{N}$ with $\vec{N}\in {N}^3$, and the energies of $M_1(\vec p)M_2(-\vec p)$    are $E^{n.i.}=E_1(p)+E_2(p)$ with $E_{1,2}(p)=(m_{1,2}^2+p^2)^{1/2}$.  The energies $E_n$ extracted from the lattice  are slightly shifted in presence of the interaction and the shift  provides rigorous information on the scattering matrix, as discussed bellow. 
In experiment, two-meson  states correspond to the two-meson decay products 
with a continuous energy spectrum. 

%The evaluation of correlation matrices including $\bar qq$ as we as $\bar qq\bar qq$ interpolating fields entail various Wick contractions, which often require quark propagators from any source to any sink location. These are not rendered by the conventional method, and  all-to-all methods such as distillation \cite{Peardon:2009gh} or stochastic distillation \cite{Morningstar:2011ka} are applied. Distillation method is  based on a particular (separable) form for smearing of quarks fields and is particularly useful for hadron spectroscopy since the spectrum $E_n$ is independent of the shape for quark smearing. 

   The mass of a hadron well below strong decay threshold is simply $m=E\vert_{\vec {p}=0}$.  The masses of resonances and near-threshold bound-states   have to be inferred from the infinite-volume scattering matrix of the one-channel (elastic) or multiple-channel (inelastic) scattering.  The bound states correspond to the poles of scattering matrix on the real axis below threshold, while the resonances masses and widths are extracted from the Breit-Wigner type fits of the corresponding cross-section or phase shift. The bound states and narrow resonances typically manifest themselves as levels that appear in addition to the expected discrete two-meson levels.  

The most widely used approach for extracting the  infinite-volume scattering matrix from the finite volume $E_n$  is based on L\"usher's seminal work and its generalizations.  In the case of elastic scattering between two hadrons with zero-total momenta, the energy $E_n=(m_1^2+p^2)^{1/2}+(m_2^2+p^2)^{1/2}$ renders momenta $p$  of each meson in the region outside the interaction. The infinite volume phase shift at that energy is given by the L\"uscher's relation $\delta_l(p)=\mathrm{atan}[\sqrt{\pi} p L/2Z_{00}(1;(pL/2\pi)^2)]$  if the partial wave $l$ dominates the scattering \cite{Luscher:1991cf}. 
This is a favorable case where one equation determines one unknown $\delta_l(E_n)$ for each energy level $E_n$.    The generalizations of this relation to multiple partial waves, non-zero total momenta, twisted boundary conditions, coupled-channel scattering and three-particle systems have also been derived in a series of papers recently.  For each energy level $E_n$ this generally leads to one (determinant) equation with several unknown $\delta_l^{a}(E_n^{cm})$ and the rigorous extraction becomes much more challenging. In this case the analysis may relay on certain parametrizations of the scattering matrix as a function of $E^{cm}$, which may render otherwise unsolvable problem tractable. It is encouraging that the Hadron Spectrum Collaboration presented  the first simulation of two-coupled channel  system $K\pi-K\eta$ and extracted the poles corresponding to strange mesons relying on the parametrization of the scattering matrix \cite{Dudek:2014qha}. 
 
 The overlaps $Z_i^n=\langle \Omega|{\cal O}_i|n\rangle$  provide wealth of information about the composition of each lattice eigenstate $|n\rangle$. This information has been used so far mostly as a qualitative guidance on the importance of various Fock components. It remains an open question how to use this rich source of information to rigorously extract physics information on the physical states (especially for smeared quarks) and analytic considerations in this direction may prove fruitful.  
 
 \section{Mesons well below threshold}
 
 Well below strong decay threshold there are no multi-hadron states, and the mass of a single hadron is extracted from $m=E\vert_{\mathbf{P}=0}$ extrapolated to $L\to \infty$, $a\to 0$ and $m_q\to m_q^{phys}$.  Particular care has to be taken concerning discretization errors related to heavy quarks and complementary methods lead to compatible results in the continuum limit. 
 
 Many precision results are available for a number of years. The continuum and chiral extrapolations of   low-lying charmonia were, for example, addressed by Fermilab/MILC \cite{Mohler:lat14} and HPQCD/MILC \cite{Galloway:lat14} collaborations recently. The resulting splittings between ground-state masses in different channels as well as spin-averaged masses of $2S$ and $1S$ charmonia  are in  good agreement with experiment.  In all simulations of charmonia and other hidden charm  channels reported here, the charm-quark annihilation contribution is omitted (while possible Wick contractions with $u/d/s$ annihilation are taken into account) and it is indeed OZI   suppressed in experiment.    The rigorous treatment of charm annihilation presents an unsolved problem due to the mixing with a number of light hadron channels and the noise in the disconnected diagrams.  
 
 The $\eta$ and $\eta'$ can strongly decay only to the three-meson states, therefore they are very narrow and can be treated using standard technique to a good approximation.  Their masses as well as the flavour mixing angle were determined as a function of $m_\pi$ by ETMC collaboration \cite{Michael:2013gka}, recovering experimental values in the chiral limit.  
 
\section{Excited mesons within a single-hadron approach}

The great majority of hadrons lie near or above strong decay threshold. Yet most of them have been treated until recently  based on a single-hadron approximation. For meson states this entails (i) using only quark-antiquark interpolating fields ${\cal O}\simeq \bar qq$ for mesons, (ii) assuming that all energy levels correspond to ``one-particle'' states  and (iii) that the mass of the  excited resonance  equals $m\!=\!E$. These are strong assumptions for the resonances, which are not asymptotic states.  The approach also ignores the effect of the threshold on near-threshold states.
 
The most extensive light isoscalar \cite{Dudek:2013yja}, $D$, $D_s$ \cite{Moir:2013ub} and $\bar cc$ \cite{Liu:2012ze} spectrum  was extracted  by the Hadron Spectrum Collaboration (HSC) on $N_f\!=\!2\!+\!1$ anisotropic configurations with $m_\pi\simeq 400~$MeV. The   mixing angle between  $(\bar uu+\bar dd)/\sqrt{2}$ and $\bar ss$ components  for isoscalar mesons mass was also calculated  and the mixing is found small for most of the states \cite{Dudek:2013yja}.   The continuum $J^{PC}$ was reliably identified using advanced spin-identification method. An impressive number of excited states was extracted in each channel with a good accuracy in spite of the disconnected contribution for isoscalars.  States  are identified with members of $\bar qq$ multiplets $nS$, $nP$, $nD$ and $nF$ based on overlaps $\langle O_i|n\rangle$, where interpolators are chosen to resemble multiplet members.  
There are several remaining states which are identified as hybrids: they do not fit $\bar qq$ multiplets and show  strong overlap with ${\cal O}\simeq \bar q F_{\mu\nu}q$.  

\section{Near-threshold  mesons (beyond  single-hadron approach)}

 Most of the exciting states found by experiments are located near thresholds, for example $X(3872)$, $Z_c^+(3900)$, $Z_b^+(10610)$, $Z_b^+(10650)$,  $D_s^0(2317)$ and $\Lambda(1405)$. The quarkonium-like states, which lie near threshold and above threshold, are listed in Tables 10 and 12 of a  review  by Brambilla et al. \cite{Brambilla:2014jmp}. Identifying whether these states arise from QCD or not, and what is their nature, presents an exciting and important challenge to the lattice community.
 
 Indeed most of the effort in the hadron spectroscopy during past few years went in going beyond the single-hadron approximation and taking into account two-hadron eigenstates rigorously. Note that majority of the studies focus on the (elastic) energy region near threshold, where the methods may be tractable at present, but one can not expect spectra of highly excited multiplets from rigorous approach soon. 
 \\
 
  \underline{$\mathbf{Z_c^+}$}:  Several charged-charmonia with quark content $\bar cc\bar d u$ were discovered recently in experiment. Most notably these are $Z^+(4430)$ with $J^P=1^+$ discovered by Belle and confirmed by LHCb, and $Z_c^+(3900)$ with unknown $J^P$ discovered by BESIII and confirmed by Belle and CLEOc  \cite{Brambilla:2014jmp}.   In is important to note that $Z_c^+(3900)$ was found in $J/\psi\,\pi$ invariant mass only through $e^+e^- \to Y(4460)\to  (J/\psi\, \pi^+)\pi^- $.  No resonant structure in $J/\psi\, \pi^+$ was seen in  $\bar B^0\to (J/\psi\, \pi^+) K^-$ by BELLE \cite{Chilikin:2014bkk}, in $\bar B^0 \to (J/\psi \pi^+) \pi^-$  by LHCb  \cite{Aaij:2014siy} or in  $\gamma p\to (J/\psi \,\pi^+) n$ by COMPASS    \cite{Adolph:2014hba}.     This might indicate that  the peak seen in $Y(4460)$ decay might not be of dynamical origin \cite{Chen:2013coa,Swanson:2014tra}.  
  
  The first search for $Z_c^+(3900)$ on the lattice considered $J/\psi\, \pi$ and $D\,\bar D^*$ scattering and no $Z_c^+$ candidate was found  \cite{Prelovsek:2013xba}. The $D\bar D^*$ scattering was  considered in   \cite{Chen:2014afa} and the authors conclude that they do not find the state either. 
  
   The most extensive lattice search for  $Z_c^+$ with mass below $4.2~$GeV in the channel  $I^G(J^{PC})=1^+(1^{+-})$ is performed in \cite{Prelovsek:2014v2}.  The major challenge is presented by the two-meson states $J/\psi\, \pi$, $\psi_{2S}\,\pi$, $\psi_{1D}\,\pi$, $D\bar D^*$, $D^*\bar D^*$, $\eta_c\,\rho$ that are  inevitably present in this channel in addition to  potential  $Z_c^+$ candidates. The spectrum of eigenstates is extracted using a number of meson-meson  and diquark-antidiquark interpolating fields. All the expected two-meson states are found but no additional candidate for $Z_c^+$ \cite{Prelovsek:2014v2}. It is  also illustrated how a simulation incorporating low-lying two-mesons states seems to render a $Z_c^+$ candidate \cite{Prelovsek:2014v1}, which is however not robust after further two-meson states around $4.2~$GeV are implemented \cite{Prelovsek:2014v2}.  It is concluded that the experimental $Z_c^+$ candidates with $I^G(J^{PC})=1^+(1^{+-})$  and a mass below $4.07~$GeV are either very broad or most likely not dominated by the $[\bar c\bar d]_{3_c}[cu]_{\bar 3_c}$ Fock component.
   \\
 
\underline{\bf $\mathbf{D_{s0}^*}$, $\mathbf{D_{s1}}$}: The quark models expected $D_{s0}^*(2317)$ and $D_{s1}(2460)$  above  $DK$ and $D^*K$ thresholds, but they were experimentally found slightly below them.  The first lattice QCD simulations that take the effect of these thresholds into account used $DK$ and $D^*K$ interpolating fields in addition to the $\bar sc$  \cite{Mohler:2013rwa,Lang:2014yfa}.  The position of thresholds is almost physical in this $N_f\!=\!2+1$ simulation with nearly physical  $m_\pi\simeq 156~$MeV. The $D^{(*)}K$ phase shift is extracted from each energy level and then parametrized in the region close to threshold using effective range formula.  The large negative scattering length is an indication for the presence of the bound states.   The effective range expansion renders the position of the poles in $S\propto (\cot\delta -i)^{-1}$ related to $D_{s0}^*(2317)$  and $D_{s1}(2460) $ close to the experimental masses.     The summary of the resulting $D_s$ spectrum for these two states as well as other $D_s$ states in summarized in Fig. 9 of \cite{Lang:2014yfa} for two values of pion masses. 
\\

\underline{$\mathbf{X(3872)}$}: A candidate for the charmonium(like) state $X(3872)$ is found $11\pm 7~$MeV below the $D\bar D^*$ threshold for $J^{PC}\!=\!1^{++}$, $I\!=\!0$, $N_f\!=\!2$ and $m_\pi\!\simeq\! 266~$MeV  \cite{Prelovsek:2013cra}. This is the first lattice simulation that establishes a candidate for $X(3872)$ in addition to $\chi_{c1}$  and the nearby scattering states $D\bar D^*$  and $J/\psi\,\omega$. The large and negative $a_0\!=\!-1.7\pm 0.4~$fm for $D\bar D^*$ scattering  is one  indication for a shallow bound state $X(3872)$.  The mass of $X$ is determined from the position of the pole in $S$ matrix which is obtained by interpolating $D\bar D^*$ scattering phase shift near threshold. The established $X(3872)$ has a large overlap with $\bar cc$ as well as $D\bar D^*$ interpolating fields \cite{Prelovsek:2013cra}. The single-hadron approach using just $\bar cc$ interpolators  renders only one level near $DD^*$ threshold just like in  previous simulations.  In this case one can not reliably establish whether this level is related to $X(3872)$ or $D(0)\bar D^*(0)$. 

In the  $I\!=\!1$ channel, only the $D\bar D^*$ and $J/\psi\,\rho$ scattering states are found, and no candidate for $X(3872)$ \cite{Prelovsek:2013cra}. This is in agreement  with a popular interpretation that $X(3872)$ is dominantly $I\!=\!0$, while its small $I\!=\!1$ component arises solely from the isospin breaking and is therefore absent in the simulation with $m_u\!=\!m_d$.  

\section{Rigorous treatment of hadronic resonances}

The rigorous treatment of a resonance in an elastic channel   $M_1M_2$ amounts to determination of the discrete spectrum including two-meson states, determination of the scattering phase shift from each energy level and making a Breit-Wigner type fit  of the phase shift as described in Section 2.  The only hadron resonance studied in this way until recently is $\pi\pi\to\rho\to \pi\pi$, which has been simulated by a number of lattice collaborations until now (see for example \cite{Dudek:2012xn}). In the following I summarize results for other channels, where only pioneering steps  have been made.  \\

\underline{$\mathbf{K^*}$, $\mathbf{\kappa}$, $\mathbf{K_0^*}$  {\bf and } $\mathbf{K_2}$}: $K^*$ mesons and in particular the $K^*(892)$ were frequently addressed in lattice simulations, but always ignoring that the $K^*(892)$  decays strongly. The simulation \cite{Prelovsek:2013ela} presents the first extraction of the  masses and widths for the $K^*$ resonances by simulating $K\pi$ scattering in $p$-wave with $I=1/2$. 
%The $K\pi$ system with total momenta $P=\tfrac{2\pi}{L}e_z~,\ \tfrac{2\pi}{L}(e_x+e_y)$ and $0$ is simulated, that allows the extraction of  phase shifts at several values of $K\pi$ relative momenta. 
A Breit-Wigner fit of the phase renders a $K^*(892)$ resonance mass $m^{lat}=891\pm 14~$MeV and the $K^*(892)\to K\pi$ coupling $g^{lat}=5.7\pm 1.6$ compared to the experimental values $m^{exp}\approx 892$ MeV and  $g^{exp}=5.72\pm 0.06$, where $g$ parametrizes the $K^*\to K\pi$ width. Mixing of $s$ and $p$-wave is taken into account 
when extracting the phase shift around the $K^*(1410)$ and $K_2^*(1430)$ resonances. 
This gives  an estimate of  the $K^*(1410)$ resonance mass $m^{lat}=1.33\pm 0.02 ~$GeV compared to $m^{exp}=1.414\pm 0.0015~$GeV assuming the experimental $K^*(1410)\to K\pi$ coupling.    

 The first simulation of two-coupled channel system $K\pi-K\eta$ was presented just around the time of this meeting \cite{Dudek:2014qha}.  
The scattering matrix in complex plane was parametrized and the parameters were extracted using the fit to the finite volume spectrum via L\"uscher-type method.  The poles corresponding $\kappa$ and $K^*(892)$ are found below $K\pi$ threshold for the employed $m_\pi\simeq 400~$MeV, while $K_0^*(1430)$ and $K_2$ are found as resonances above threshold.  
\\

\underline{$\mathbf{D_{0}^*}$ {\bf and} $\mathbf{D_{1}}$}:  The first rigorous simulation of a hadronic resonance that contains charm quarks addresses the  broad scalar $D_0^*(2400)$ and the axial $D_1(2430)$ charmed-light mesons, which appear in    $D\pi$ and $D^*\pi$ scattering   \cite{Mohler:2012na}. The simulation is done for $N_f=2$ and $m_\pi\simeq 266~$MeV. The resonance parameters are obtained using a Breit-Wigner fit to the elastic phase shifts. The resulting $D_0^*(2400)$  mass is $351\pm 21~$MeV above the spin-average $\tfrac{1}{4}(m_D+3m_{D^*})$, in agreement with the experimental value of $347\pm 29~$MeV above. The resulting $D_0^*\to D\pi$ coupling $g^{lat}=2.55\pm 0.21~$GeV is close to the experimental value $g^{exp}\le1.92\pm 0.14~$GeV, where $g$ parametrizes the width $\Gamma\equiv g^2p^*/s$. The results for $D_1(2430)$ are also found close to the experimental values; these are obtained by appealing to the heavy quark limit, where the neighboring resonance $D_1(2420)$ is narrow.

The charmed scalar meson puzzle wonders why the strange $D_{s0}^*(2317)$ and the non-strange $D_0^*(2400)$ charmed scalar mesons have a mass within 1 MeV of each other experimentally, while one would naively expect a larger slitting $O(m_s)$. The question is whether this near degeneracy is due to the strange meson being pushed down or the non-strange one being pushed up. This puzzle can be addressed by considering the lattice results for $D_0^*(2400)$  \cite{Mohler:2012na}, which is found as a resonance in $D\pi$, and $D_{s0}^*(2317)$ \cite{Mohler:2013rwa}, which is found as a pole below $DK$.  Both masses are found close to experiment. This favors the interpretation that the near degeneracy is a consequence of strange meson being pushed down due to $DK$ threshold. On the other hand, the interpretation that $D_0^*(2400)$ is pushed up due to tetra quark Fock component $\bar u\bar ss c$  is disfavored since $N_f=2$ simulation \cite{Mohler:2012na} renders its mass close to the experiment without any strange content in valence or sea sectors. \\

\underline{$\mathbf{a_1}$ {\bf and} $\mathbf{b_1}$}: The light axial-vector resonances $a_1(1260)$ and $b_1(1235)$ are explored for $N_f\!=\!2$  by simulating the corresponding scattering channels $\rho \pi$ and $\omega\pi$ \cite{Lang:2014tia}. 
Interpolating fields $\bar qq$ and  $\rho\pi$ or $\omega\pi$ are used to extract the $s$-wave phase shifts for the first time. It is  assumed that $\rho$ and $\omega$ are stable, which is  justified  in the energy region of interest for the employed  parameters $m_\pi\simeq 266~$MeV and $L\simeq 2~$fm.
A Breit-Wigner fit  of the phase shift gives the $a_1(1260)$ resonance mass  $m_{a_1}^{\textrm{res}}=1.435(53)(^{+0}_{-109})~$GeV compared to $m_{a_1}^{\textrm{exp}}=1.230(40)~$GeV. The $a_1$ width $\Gamma_{a_1}(s)\equiv g^2 p/s$ is parametrized in terms of the coupling, which results in  $g_{a_1\rho\pi}=1.71(39)~$GeV compared to  $g_{a_1\rho\pi}^{\textrm{exp}}=1.35(30)~$GeV derived from experiment.  

\section{Conclusions}

I have reviewed recent lattice results for conventional and exotic mesons.  A number of precise results for states well below strong decay threshold are available for a number of years and there is impressive agreement with experiment.  I reported on the first   rigorous simulations aimed at near-threshold states $Z_c^+$, $X(3872)$, $D_{s0}^*(2317)$ as well as light, strange and charmed resonances. 
 
 \begin{acknowledgement}
 I would like to thank  C.B. Lang, L. Leskovec, D. Mohler and R. Woloshyn for the pleasure of collaborating on the described topics. This work is supported by ARRS project number N1-0020 and FWF project number I1313-N27. 

\end{acknowledgement}

\bibliographystyle{h-physrev4}
\bibliography{Lgt}

\end{document}